\documentclass{LMCS}

\def\dOi{9(3:20)2013}
\lmcsheading%
{\dOi}
{1--19}
{}
{}
{Mar.~16, 2013}
{Sep.~17, 2013}
{}

\usepackage{enumerate}
\usepackage{hyperref}
\usepackage{amssymb}

\theoremstyle{plain}
\newtheorem{theorem}[thm]{Theorem}
 \newtheorem{lemma}[thm]{Lemma}

\begin{document}

\title[Induction in Algebra: a First Case Study]{Induction in Algebra:
  a First Case Study\rsuper*}

\titlecomment{{\lsuper*}This is a revised and extended journal version
  of the author's LICS 2012 conference paper \cite{LICS2012}.}

\author[P.~Schuster]{Peter Schuster}
\address{Pure Mathematics, University of Leeds, Leeds LS2 9JT, England}
\email{pschust@maths.leeds.ac.uk}

\ACMCCS{[{\bf Theory of computation}]: Logic---Proof
  theory\,/\,Constructive mathematics}


\begin{abstract}
Many a concrete theorem of abstract algebra admits a short and elegant proof
by contradiction but with Zorn's Lemma (ZL). A few of these theorems have
recently turned out to follow in a direct and elementary way from the
Principle of Open Induction distinguished by Raoult. The ideal objects
characteristic of any invocation of ZL are eliminated, and it is made
possible to pass from classical to intuitionistic logic. If the theorem has
finite input data, then a finite partial order carries the required instance
of induction, which thus is constructively provable. A typical example is
the well-known theorem ``every nonconstant coefficient of an invertible
polynomial is nilpotent''.
\end{abstract}

\maketitle

\section{Introduction}

Many a concrete theorem of abstract algebra admits a short and elegant proof
by contradiction but with Zorn's Lemma (ZL). A few of these theorems have
recently turned out to follow in a direct and elementary way from the
Principle of Open Induction (OI) distinguished by Raoult \cite{raoult:open}.
A proof of the latter kind may be extracted from a proof of the former sort.
If the theorem has finite input data, then a finite partial order carries
the required instance of induction, which thus is provable by mathematical
induction---or, if the size of the data is fixed, by fully first-order
methods.

But what is Open Induction? In a nutshell, OI is transfinite induction for
subsets of a directed-complete partial order that are open with respect to
the Scott topology. While OI was established \cite{raoult:open} as a
consequence of ZL,\ by complementation these two principles are actually
equivalent \cite{hub:ind} with classical logic but in a natural way. Hence
OI is the fragment of transfinite induction of which the corresponding
minimum principle just is ZL.

Our approach is intended as a contribution to a partial realisation in
algebra \cite{coq:logicalgebra} of the revised Hilbert Programme \`{a} la
Kreisel and Feferman (see \cite{cro:fin} for a recent account including
references), and was motivated by related work in infinite combinatorics 
\cite{berger:open,coq:combinatorics,coqper:groebner,raoult:open} as well as
by the methods of dynamical algebra \cite%
{cos:dyn,perdry:noether,yengui:maximal} and formal topology \cite%
{lombardi:algdyn,sam:ifs,sam:som}. In Hilbert's terminology, the
\textquotedblleft ideal objects\textquotedblright\ characteristic of any
invocation of ZL are eliminated by passing to OI, and it is made possible to
work with \textquotedblleft finite methods\textquotedblright\ only, e.g.~to
pass from classical to intuitionistic logic.

A typical example, studied before \cite{per:spe,rich:trivial} and taken up
in this paper, is the well-known theorem \textquotedblleft every nonconstant
coefficient of an invertible polynomial is nilpotent". More formally, this
can be put as%
\begin{equation}
fg=1\rightarrow \exists e\left(u^{e}=0\right)   \label{nc}
\end{equation}%
where $f$ and $g$ are polynomials with coefficients in an arbitrary commutative ring%
\[
f=\sum_{i=0}^{n}a_{i}T^{i}\,,\quad g=\sum_{i=0}^{m}b_{j}T^{j}
\]%
and $u= a_{i_{0}}$ where $1\leq i_{0}\leq n$. The customary short and elegant proof of (\ref{nc}) works by reduction to the case of polynomials over an integral domain 
\begin{equation*}
fg=1\rightarrow u=0  
\end{equation*} 
or, equivalently, by reduction modulo any prime ideal $P$ of the given ring: 
\begin{equation}
fg=1\rightarrow \forall P\left( u\in P\right) \,.  \label{mid}
\end{equation}%
This special case is readily settled by looking at the degrees, or more explicitly by a polynomial trick due to Gau\ss\ \cite{coq:prague,lom:int}. In order to reduce (\ref{nc})  to (\ref{mid}),  it is natural to invoke%
\begin{equation}
\forall P\left( u\in P\right) \rightarrow \exists e\left(
u^{e}=0\right) \,.  \label{kl}
\end{equation}%
But the latter, a variant of Krull's Lemma, is normally deduced from ZL by
a proof by contradiction, which is anything but an argument using only
finite methods. In addition, a universal quantification
over prime ideals $P$ occurs, which are ideal objects (see e.g.~%
\cite{cro:fin}).

These foundational issues aside, there is a practical problem. By
decomposing (\ref{nc}) into (\ref{mid}) and (\ref{kl}) one virtually loses
the computational information the hypothesis of (\ref{nc}) is made of;\ in
particular \cite{per:spe,rich:trivial} the proof falls short of being an
algorithm for computing an exponent $e$ under which the nilpotent $u$ 
vanishes. However, we can still extract a proof that is based on induction
over a finite partial order;\ and the proof tree one can grow alongside the
induction encodes an algorithm which computes the desired exponent.

That our method does work may seem less surprising if one takes into account
that the theorem has already seen constructive proofs before \cite%
{per:spe,rich:trivial}, and that an entirely down-to-earth proof is possible
anyway \cite[Chapter 1, Exercise 2]{AMD}. Needless to say, each of those
proofs embodies an algorithm;\ one of them \cite{per:spe} has even been
partially implemented in Agda, a proof assistant based on Martin--L\"{o}f
type theory.

Just as the proof in \cite{per:spe}, our constructive proof is gained from a
given classical one, the one by reduction to the case of integral domains we
have mentioned above. As compared with \cite{per:spe}, we keep somewhat
closer to the classical proof. The price we have to pay is that we have to
suppose certain decidability hypotheses, which need to---and can---be
eliminated afterwards by a variant of the G\"{o}del--Gentzen and
Dragalin--Friedman translations. In \cite{per:spe} a simpler instance of
this elimination method is built directly into the proof.

To be slightly more specific, we we first turn the indirect proof of (\ref%
{kl})\ with ZL into a direct deduction from OI;\ and then transform the
latter into a constructive proof of (\ref{nc}) by induction over a finite
poset. This is possible because the hypothesis of (\ref{nc})---unlike the
one of (\ref{kl})---consists of computationally relevant information about a
finite amount of elementary data: of nothing but the finitely many equations 
\[
a_{0}b_{0}=1,\,a_{0}b_{1}+a_{1}b_{0}=0\,,\ldots ,\,a_{n}b_{m}=0\,.
\]%
Heuristics aside, all this makes redundant the reduction, and the prime
ideals disappear.

\subsection{Preliminaries}

\subsubsection{Foundations}

The overall framework of this note is \emph{constructive} algebra \`{a} la
Kronecker and Bishop \cite{lomqui:pmbook,min:bib}. Due to the corresponding
choice of intuitionistic logic, one or the other assumption needs to be made
explicit that would be automatic in \emph{classical} algebra, by which we
mean algebra as carried out within \textbf{ZFC} set theory and thus, in
particular, with classical logic. For example, we say that an assertion $A$
is \emph{decidable} whenever $A\vee \lnot A$ holds; and that a subset $S$ of
a set $T$ is \emph{detachable }if $t\in S$ is decidable for each $t\in T$. 

As moreover the principle of countable choice will not occur, let alone the
one of dependent choice, our constructive reasoning can be carried out
within (a suitable elementary fragment of) the Constructive
Zermelo--Fraenkel Set Theory \textbf{CZF} which Aczel \cite%
{AczelInter,AczelChoice,AczelInd} has interpreted within Martin-L{\"{o}}f's 
\cite{ITT} Intuitionistic Theory of Types. Unlike Friedman's \cite%
{friedman:foundations} impredicative Intuitionistic Zermelo--Fraenkel Set
Theory \textbf{IZF}, this \textbf{CZF }does not contain the axiom of power
set. Hence in \textbf{CZF }an unrestricted quantification over
subsets---such as the one crucial for this paper, over all prime ideals of
an arbitrary ring---in general is a quantification over the members of a
class.

\subsubsection{Rings}

Throughout this paper, $R$ will denote a commutative ring (with unit). We
briefly recall some related concepts \cite{AMD}. An \emph{ideal }of $R$ is a
subset $I$ that contains 0, is closed under addition, and satisfies%
\[
s\in I\rightarrow rs\in I 
\]%
for all $r,s\in R$. We write $(S)$ for the ideal \emph{generated} by a
subset $S$ of $R$: that is, $(S)$ consists of the linear combinations $%
r_{1}s_{1}+\ldots +r_{n}s_{n}$ of elements $s_{1},\ldots ,s_{n}$ of $S$ with
coefficients $r_{1},\ldots ,r_{n}$ from $R$.

A \emph{radical ideal }of $R$ is an ideal $I$ such that%
\[
r^{2}\in I\rightarrow r\in I 
\]%
for all $r\in R$. The \emph{radical} 
\[
\sqrt{I}=\{r\in R:\exists e\in \mathbb{N}\,(r^{e}\in I)\} 
\]%
of an ideal $I$ is a radical ideal with $I\subseteq \sqrt{I}$. An ideal $I\ $%
is a radical ideal if and only if $I=\sqrt{I}$. The radical $\sqrt{0}$ of
the zero ideal $0=\{0\}$ is the \emph{nilradical}, and its elements are the 
\emph{nilpotents}.

An ideal $P$ is a \emph{prime ideal }if $1\notin P$ and%
\begin{equation}
ab\in P\rightarrow a\in P\vee b\in P  \label{prime}
\end{equation}%
for all $a,b\in R$. Clearly, every prime ideal is a radical ideal. A ring $R$
is an \emph{integral} \emph{domain}---for short, a \emph{domain}---if $1\neq
0$ in $R$ and 
\begin{equation}
ab=0\rightarrow a=0\vee b=0  \label{domain}
\end{equation}%
for all $a,b\in R$. A quotient ring $R/P$ is a domain if and only if $P$ is
a prime ideal.

\subsubsection{Induction}

Let $(X,\leq )$ be a partial order. We do not specify from the outset
whether $X$ is a set in the sense of \textbf{CZF}, which in the case of our
definite interest will anyway be the case, but during heuristics will depend
on the choice of a more generous set theory such as \textbf{IZF}. 
Unless specified otherwise every quantification over the variables $x$, $%
x^{\prime }$, $y$, and $z$ is understood as over the elements of the partial
order $X$ under consideration.

Let $U$ be a predicate on $X$. We say that $U$ is \emph{progressive }if%
\begin{equation}
\forall x\,(\forall y>x\,U(y)\rightarrow U(x)),  \label{pro}
\end{equation}%
where $y>x$ is understood as the conjunction of $y\geq x$ and $y\neq x$.
About the antecedent of (\ref{pro}), note that $y>x$ is used rather than $y<x
$, as is common in other contexts;\ our choice allows us to avoid reversing
the naturally given order (i.e., inclusion) later on. Also, in the relevant
instantiations below, the predicate $U$ will define---and be identified
with---a subset of the set $X$; and $\leq $ will be a decidable relation
(that is, for all $x,y\in X$, the assertion $x\leq y$ is decidable).

By \emph{induction for }$U$ \emph{and }$X$ we mean the following:

\begin{center}
\emph{If }$U$\emph{\ is progressive, then }$\forall x\,U( x) $\emph{.}
\end{center}

\noindent%
Classically, induction holds for \emph{every} $U$ precisely when $X$ is 
\emph{well-founded} in the sense that every inhabited predicate on $X$ has a
maximal element---or, in classically equivalent terms, that there is no
strictly increasing sequence in $X$.

We will use induction in cases in which $X$ has a least element $\bot $, in
which $U( \bot ) $ is equivalent to $\forall x\,U( x) $ whenever $U$ is 
\emph{monotone: }that is, if $x\leq y$, then $U( x) $ implies $U( y) $. Note
finally that if $U$ is progressive, then $U$ is satisfied by every maximal
element of $X$, and thus by the greatest element $\top $ of $X$ whenever
this exists.

\section{Noetherian Rings}

As a warm-up we first revisit the perhaps historically first---albeit
implicit---occurrence of induction in algebra: Krull's proof \cite[pp.~8--9]%
{krull:ideal} of the Lasker--Noether decomposition theorem for Noetherian
rings. According to one of the constructively meaningful variants of this
concept \cite{jacobsson:standard}, a ring is \emph{Noetherian }if induction
holds for the partial order consisting of the finitely generated ideals,
which by the way is a set in \textbf{CZF}. We now prove, using this instance
of induction, the following corollary of the \emph{Lasker--Noether Theorem}:

\begin{description}
\item[LN] \emph{The radical }$\sqrt{I}$ \emph{of every finitely generated
ideal }$I$\emph{\ of a commutative Noetherian ring\ is the intersection of
finitely many finitely generated prime ideals.}
\end{description}

\noindent Constructive proofs given before \cite{perdry:noether} with
related notions of \textquotedblleft Noetherian\textquotedblright\ have
motivated our choice of this example; see also \cite{perdry:noeord}.

Before proving LN\ we recall a well-known fact (see e.g.~the proof of \cite[%
Proposition 1.8]{AMD}), which however will be crucial for a large part of
this paper.

\begin{lemma}
\label{key}Let $R$ be a commutative ring. If $I$ is an ideal of $R$, and $%
a,b\in R$, then%
\[
\sqrt{I+Ra}\cap \sqrt{I+Rb}=\sqrt{I+Rab}. 
\]
\end{lemma}

\proof
Since $\supseteq $ is clear, we only verify $\subseteq $. Let $x\in \sqrt{%
I+Ra}$ and $x\in \sqrt{I+Rb}$, which is to say that $x^{k}=u+sa$ and $x^{\ell }=v+tb
$ where $k,\ell \in \mathbb{N}$, $u,v\in I$, and $s,t\in R$. Then%
\[
x^{k}x^{\ell }=\underbrace{uv+utb+sav}_{\in I}+st\,ab
\]%
and thus $x^{k+\ell }\in I+Rab$ as required.
\qed

In addition to this, and the aforementioned instance of induction, we need
to employ a distinction-by-cases that is known as \emph{Strong Primality
Test (SPT)} \cite{perdry:noether}. This says that for every finitely
generated ideal $I$ of $R$ one of the following three conditions is
fulfilled:

\begin{enumerate}
\item[(i)] $I=R$, which is to say that $1\in I$;

\item[(ii)] for all $a,b\in R$, if $ab\in I$, then either $a\in I$ or $b\in
I $;

\item[(iii)] there are $a,b\in R$ for which $ab\in I$ but neither $a\in I$
nor $b\in I$.
\end{enumerate}

\noindent In other words, the SPT\ tells us whether $I\ $is a prime ideal;
and moreover if the answer is in the negative, then the SPT\ provides us
with witnesses for this fact. Clearly SPT is classically valid, but it also
holds constructively whenever $R$ is a fully Lasker--Noether ring 
\cite{perdry:noether}.

To prove LN\ by induction, consider \textquotedblleft $\sqrt{I}$ is the
intersection of finitely many prime ideals\textquotedblright\ as a predicate 
$U$ of the finitely generated ideals $I$ of $R$. To show that $U$ is
progressive, let $I$ be a finitely generated ideal of $R$. If $I=R$, then
clearly $U(I)$; if $I$ is a prime ideal, then in particular $\sqrt{I}=I$,
and thus $U(I)$. If however $a,b\in R$ are as in case (iii) of SPT, then 
$I\subsetneqq I+Ra$ and $I\subsetneqq I+Rb$ but $I=I+Rab$; whence $U(I+Ra)$
and $U(I+Rb)$ hold by induction, and $U(I)$ follows with Lemma \ref{key} at
hand.

This proof of LN can be carried over in a relatively easy way to the full
Lasker--Noether theorem, and as such can be viewed as an \textquotedblleft
unwinding\textquotedblright\ not only of Krull's proof, but also of the
better-explicated proofs given in \cite{AMD,north:ideal}. We refrain from
doing this transfer to the full theorem, for no further insight into the
method would be gained.

\section{Some Induction Principles}

When does induction hold more in general, regardless of the specific partial
order under consideration? A fairly general induction principle has been
coined by Raoult \cite{raoult:open} as follows. A partial order $X$ is \emph{%
chain-complete} if {every chain }$Y$ in $X$ has a least upper bound $\bigvee
Y\in X$. {A predicate $U$ on $X$ is \emph{open} in the lower topology if,
for every chain }$Y$ in $X$, {\ 
\[
U(\bigvee Y)\rightarrow \exists x\in Y\,U(x). 
\]%
(Think of the elements }$x$ of $Y$ as of \textquotedblleft
neighbourhoods\textquotedblright\ of the \textquotedblleft
limit\textquotedblright\ $\bigvee Y$ of $Y$.) Now Raoult's \emph{Open
Induction (OI)}{\ is induction for chain-complete }$X$ and open $U$. It is
easy to see \cite{raoult:open} that OI follows classically from \emph{Zorn's
Lemma (ZL)}, and thus holds in \textbf{ZFC};\ moreover OI\ and ZL\ are
classically equivalent \cite{hub:ind}, by complementation.

Open Induction implies \emph{Well-Founded Induction (WI)} which is induction
for well-founded $X$ and arbitrary $U$.\footnote{%
This principle is also known as \emph{Noetherian Induction} and, in the case
of a well-ordered $X$, as \emph{Transfinite Induction}; see e.g.~\cite[p.~21]%
{Cohn:Univ}.} In fact, if $X$ is well-founded, then every chain in $X$ has a
greatest element; whence $X$ is chain-complete, and every $U$ is open.
Unlike OI, WI is provable in \textbf{ZF}, but most partial orders that are
classically well-founded lack this property from a constructive perspective.
For the notion of a Noetherian ring one can, as we have recalled above,
circumvent this problem by simply defining a commutative ring to be
Noetherian if one can perform induction on the finitely generated ideals 
\cite{jacobsson:standard}.

We say that a partial order $X$ is \emph{finite} if $X$ has finitely many
elements (that is, $X=$ $\{x${{$_{1},\dots ,x_{n}\}$ for some} }$n\geq 0$,
which includes the case $n=0$ of $X=\emptyset $), and if, in addition, $\leq 
$ is a decidable relation. In this case, $X$ is a \emph{discrete} set, which
is to say that equality $=$ is decidable;\footnote{%
For any such $X$, in particular, there is no need to distinguish between
\textquotedblleft $X$ is finite\textquotedblright\ and \textquotedblleft $X$
is finitely enumerable\textquotedblright , as is customary in constructive
mathematics: these two variants of the notion of a finite set coincide in
the case of a discrete set \cite[p.~11]{min:bib}.}\ whence so is $<$ too.

Classically, every finite $X$ is well-founded; whence WI implies \emph{%
Finite Induction (FI)}: that is, induction for finite $X$ and for arbitrary $%
U$. Unlike WI, this FI is even constructively provable, as in \textbf{CZF},
by means of mathematical induction. To see this note first that if $X$ is
finite, then one can exhibit a maximal element $x$ of $X$, for which $%
U\left( x\right) $ anyway. In fact, if $X$ has only finitely many elements,
then $\lnot \forall x\exists y\left( x<y\right) $, which is to say that $%
\exists x\lnot \exists y\left( x<y\right) $ if, in addition, $\leq $ is
decidable.

\section{A Proof Pattern}

In all cases considered later in this paper, $X$ consists in certain ideals
of a commutative ring, with the partial order given by inclusion, for which $%
\wedge $ simply is $\cap $. Following the terminology which is standard for
this special case, we also say for a general partial order $X$ that $x\in X$
is \emph{reducible }if there are $y,z\in X$ such that $x<y$, $x<z$, and $%
x=y\wedge z$. Here $x=y\wedge z$ is to be understood as that $x$ is the
greatest lower bound of $y$ and $z$: that is, 
\[
\forall x^{\prime }\,(x^{\prime }\leq x\longleftrightarrow x^{\prime }\leq
y\wedge x^{\prime }\leq z). 
\]%
In the following, let again $U$ be a predicate on a partial order $X$. We
say that $U$ is \emph{good }if, for every $x\in X$, either $U(x)$ or $x$ is
reducible. Also, we say that $U$ is\emph{\ meet-closed }whenever if $%
x=y\wedge z$ in $X$, then $U(x)$ follows from $U(y)\wedge U(z)$.

\begin{lemma}
Let $U$ be a predicate on a partial order $X$. If $U$ is meet-closed and
good, then $U$ is progressive.
\end{lemma}

All this allows us to state a proof pattern that has been prompted by \cite%
{CiE2012}:

\begin{theorem}
\label{PP}Assume that induction holds for $U$ and $X$. If $U$ is meet-closed
and good, then $\forall x\,U( x) $.
\end{theorem}

We will next look into applications of this proof pattern.

\section{Krull's Lemma with Open Induction}

Let $R$ again be a commutative ring. For heuristic purposes we first look at
the contrapositive of a variant of \emph{Krull's Lemma}:

\begin{description}
\item[KL] \emph{If~}$r\in P$\emph{~for\ all\ prime~ideals~}$P$\emph{~of~}$R$%
\emph{,\ then~}$r\in \sqrt{0}$\emph{.}
\end{description}

\noindent%
As is well known (see, for example, the proof of \cite[Proposition 1.8]{AMD}%
), with ZL at hand one can give a proof by contradiction of KL: if $%
r^{e}\neq 0$ for all $e\in \mathbb{N}$, which is to say that $0\notin S$ for
the multiplicative set $S=\{r^{e}:e\in \mathbb{N}\}$, then by ZL there is a
prime ideal $P$ of $R$ with $P\cap S=\emptyset $ and, in particular, $%
r\notin P$. If $R$ is Noetherian in the sense of \cite{jacobsson:standard},
then KL is an instance of LN, which we have already reproved by induction,
without any talk of ZL.

For an arbitrary ring $R$, KL can be deduced from OI in a direct way, by
Theorem \ref{PP} and as follows. As OI requires a chain-complete $X$, this
time we have to let $X$ consist of all the radical ideals of $R$. This $X$
actually is a frame with $\bot =\sqrt{0}$ and $\top =R$, and a set in 
\textbf{IZF}. Accordingly, we need a strong primality test for arbitrary
(radical) ideals, but remember that we are still doing heuristics.

Now, let~$r\in P$~for\ all\ prime~ideals~$P$~of~$R$. To prove $r\in \sqrt{0}$
we define the predicate $U$ on $X$ by $U(F)\equiv r\in F$ whenever $F\in X$,
for which clearly $U(\top )$. Further, $U$ is meet-closed and monotone. In
particular, to show that $U(F)$ holds for all $F\in X$ is tantamount to
showing that $U(\bot )$, i.e.~$r\in \sqrt{0}$, which is exactly what we are
after.

To see that $U$ is good, let $F\in X$:\ that is, $F=$ $\sqrt{I}$ for some
ideal $I$ of $R$. If $F=R$, then trivially $U(F)$; if $F$ is a prime ideal,
then $U( F) $ by hypothesis; if however there are $a,b\in R$ such that $%
ab\in F$ but neither $a\in F$ nor $b\in F$, then $F\subsetneqq \sqrt{I+Ra}$
and $F\subsetneqq \sqrt{I+Rb}$ but $\sqrt{I+Rab}=F$; whence $F$ is reducible
by Lemma \ref{key}. In all, Theorem \ref{PP} applies.

\section{Nilpotent Coefficients with Finite Induction}

We now can proceed to our principal example. Once more let $R$ be a
commutative ring, which we now suppose to be a set in \textbf{CZF}. Recall
that $r\in R$ is said to be a \emph{unit }or \emph{invertible }if there is $%
s\in R$ such that $rs=1$. As usual let $R[T]$ stand for the ring of
polynomials with indeterminate $T$ and coefficients from $R$. Pick an
arbitrary $f\in R[T]$ and write it as $f=\sum_{i=0}^{n}a_{i}T^{i}$. We
consider the following statement about \emph{nilpotent coefficients}:

\begin{description}
\item[NC] \emph{If }$f$\emph{\ is a unit of }$R[T]$\emph{, then }$a_{i}\in 
\sqrt{0}$ \emph{for }$i>0$\emph{.}
\end{description}

\noindent%
This is well-known, and can be put as 
\[
\exists g\in R[T]\,\left( fg=1\right) \rightarrow \forall i\in \{1,\ldots
,n\}\exists e\in \mathbb{N\,}\left( a_{i}^{e}=0\right) 
\]%
or equivalently as 
\[
\forall g\in R[T]\,\forall i\in \{1,\ldots ,n\}\left( fg=1\rightarrow
\exists e\in \mathbb{N\,}\left( a_{i}^{e}=0\right) \,\right) \,. 
\]%
Pick $g\in R[T]$ and $i_{0}\in \{1,\ldots ,n\}$, and set $u=a_{i_{0}}$.
Hence the essence of NC is%
\[
fg=1\rightarrow \exists e\in \mathbb{N\,}\left( u^{e}=0\right) \,. 
\]%
Write $f$, $g$ as%
\[
f=\sum_{i=0}^{n}a_{i}T^{i}\,,\quad g=\sum_{i=0}^{m}b_{j}T^{j}\,\,. 
\]%
Since then 
\[
fg=\sum_{k=0}^{n+m}c_{k}T^{k}\,,\text{\quad }c_{k}=\sum_{i+j=k}a_{i}b_{j}\,, 
\]%
the hypothesis $fg=1$ can be expressed as 
\begin{equation}
c_{0}=1\wedge c_{1}=0\wedge \ldots \wedge c_{n+m}=0\,  \label{implizit}
\end{equation}%
or even more explicitly as%
\begin{equation}
a_{0}b_{0}=1\wedge a_{0}b_{1}+a_{1}b_{0}=0\wedge \ldots \wedge
a_{n}b_{m}=0\,.  \label{explizit}
\end{equation}%
In particular $fg=1$ is a finite conjunction of atomic formulas of the
language of rings.

By swapping $f$ and $g$ one could also take $u$ from the $b_{1},\ldots
,b_{m} $ rather than from the $a_{1},\ldots ,a_{n}$. We do not follow this
option, but note for later use that under the hypothesis $fg=1$ we have%
\begin{equation}
\forall i>0\,\left( a_{i}\in F\right) \leftrightarrow \forall j>0\,\left(
b_{j}\in F\right)  \label{invneu}
\end{equation}%
for every ideal $F$ of $R$;\ in the particular case $F=0$ this means%
\begin{equation}
\forall i>0\,\left( a_{i}=0\right) \leftrightarrow \forall j>0\,\left(
b_{j}=0\right) \,.  \label{invorg}
\end{equation}%
As for (\ref{invneu}), let $i\in \{1,\ldots ,n\}$. If $b_{j}\in F$ for all $%
j\in \{1,\ldots ,\min \{i,m\}\}$, then 
\[
F\ni c_{i}=\underbrace{a_{0}b_{i}+\ldots +a_{i-1}b_{1}}_{\in F}+a_{i}b_{0} 
\]%
by (\ref{implizit}) and thus $a_{i}b_{0}\in F$, from which we get $a_{i}\in
F $ because $a_{0}b_{0}=1$ by (\ref{explizit}). A similar argument deals
with the converse implication in (\ref{invneu}).

As Richman has observed \cite{rich:trivial}, the statement NC above

\begin{quote}
\ldots\ admits an elegant proof upon observing that each $a_{i}$ with $i\geq
1$ must be in every prime ideal of $R$, and that the intersection of the
prime ideals of $R$ consists of the nilpotent elements of $R$. This proof
gives no clue as to how to calculate $n$ such that $a_{i}^{n}=0$, while such
a calculation can be extracted from the proof that we present.
\end{quote}

Richman's fairly short proof \cite{rich:trivial} is in fact a clever
\textquotedblleft nontrivial use of trivial rings\textquotedblright , and of
course is fully constructive. The elementary character of NC anyway suggests
an equally elementary proof, by mathematical induction, as indicated in \cite%
[Chapter 1, Exercise 2]{AMD}. 
Just as for the approach \cite{per:spe} via point-free topology, the point
of our subsequent considerations is that we \textquotedblleft
unwind\textquotedblright\ the classical proof that Richman has rightly
deemed \textquotedblleft elegant\textquotedblright , and thus get a
constructive one from which the required exponent can equally be extracted.

\subsection{A Classical Proof with Krull's Lemma}

We next review the \textquotedblleft elegant\textquotedblright\ proof of NC
which works by reduction to the case of a domain.\emph{\ }Since in a domain
every nilpotent is zero, NC for domains reads as

\begin{description}
\item[NC$_{\text{int}}$] \emph{Let }$R$ \emph{be a domain. If }$f$ \emph{is
a unit of }$R[T]$\emph{, then }$a_{i}=0$\emph{\ for }$i>0$\emph{.}
\end{description}

\noindent%
With the notation from before, the essence of NC$_{\text{int}}$ is%
\[
fg=1\rightarrow u=0\,. 
\]%
A quick proof goes as follows. Let $R$ be a domain. Then the degree of
non-zero polynomials (remember that we are in a classical setting) satisfies%
\begin{equation}
\deg \left( fg\right) =\deg \left( f\right) +\deg \left( g\right) \,.
\label{degful}
\end{equation}%
Now if $fg=1$, then $\deg \left( fg\right) =0$ in view of (\ref{implizit})
and of $1$ $\neq 0$ (recall that $R$ is a domain); whence $\deg \left(
f\right) =0$ and thus in particular $u=0$\emph{\ }as required. Note that the
special case 
\begin{equation}
\deg \left( fg\right) =0\rightarrow \deg \left( f\right) =0  \label{deghal}
\end{equation}%
of (\ref{degful}) is sufficient for proving NC$_{\text{int}}$.

Alternatively one can prove NC$_{\text{int}}$ by means of a trick that has
been ascribed to Gau\ss\ \cite{coq:prague,lom:int}, and which is nothing but
an explicit version of (\ref{deghal}). To this end let again $R$ be a
domain, and suppose that $fg=1$. Now assume towards a contradiction that $%
a_{i}\neq 0$ for some $i>0$; whence by (\ref{invorg}) also $b_{j}\neq 0$ for
some $j>0$. Pick $i,j$ both maximal with these properties, for which 
\begin{equation}
0=c_{i+j}=\underbrace{\sum_{q>j}a_{p}b_{q}}_{=0}+a_{i}b_{j}+\underbrace{%
\sum_{p>i}a_{p}b_{q}}_{=0}  \label{GaussOrg}
\end{equation}%
and thus $a_{i}b_{j}=0$; whence either $a_{i}=0$ or $b_{j}=0$, a
contradiction.

Following a time-honoured tradition, the case of NC for an arbitrary ring $R$
is handled by working modulo a generic prime ideal $P$ of $R$, for which the
quotient ring $R/P$ is indeed a domain. Hence if $P$ is a prime ideal, then
we can apply NC$_{\text{int}}$ with $R/P$ in place of $R$. This yields that
for all prime ideals $P$ of $R$ we have $u=0$ in $R/P$, which is to say that$%
~u\in P$. In all, $u$ is nilpotent by KL, which we have deduced from OI\
before.

\subsection{Discussion and Outline\label{discuss}}

In the classical proof above one first aims at the implication%
\[
fg=1\rightarrow \forall P\left( u\in P\right) \,, 
\]%
and then combines it with the appropriate instance of KL:%
\[
\forall P\left( u\in P\right) \rightarrow \exists e\left( u^{e}=0\right) \,. 
\]%
The corresponding invocation of ZL or OI aside, there is another
foundational problem with this classical proof: it rests upon a universal
quantification over all the prime ideals of $R$, which are ideal objects in
Hilbert's sense. This is reflected by the practical problem that the
computational information of $fg=1$ is virtually lost when passing to $%
\forall P\left( u\in P\right) $.

However, in the given situation one can do better. Before following our own
route, we briefly sketch the dual of the translation \cite{per:spe} of the
\textquotedblleft elegant\textquotedblright\ proof into point-free terms.%
\footnote{%
This has kindly been pointed out to us by one of the anonymous referees.}
The key move is to rewrite the classical proof by reduction to $R/P$ where $P
$ is any prime ideal, by replacing every occurrence of $x\in P$ by one of $%
D\left( x\right) =0$. Here one considers the bounded distributive lattice 
\cite{joyal:zar}---see also, for instance, \cite{ban:rad,joh:sto}---that is
generated by the symbolic expressions $D\left( x\right) $ indexed by the $%
x\in R$ and subject to the relations 
\[
\begin{array}{cc}
D(1)=1\,\text{,} & D(xy)=D(x)\wedge D(y)\,\text{,} \\ 
D(0)=0\,\text{,} & D(x+y)\leq D(x)\vee D(y)\,\text{,}%
\end{array}%
\]%
which are dual to the characteristic properties of a prime ideal $P$:%
\[
\begin{array}{cc}
1\not\in P\mathfrak{\,}, & xy\in P\leftrightarrow x\in P\vee b\in P\mathfrak{%
\,}, \\ 
0\in \mathfrak{\,}P\,, & x\in P\wedge y\in P\rightarrow x+y\in P\,.%
\end{array}%
\]%
Having shown by rewriting that $D\left( u\right) =0$, the key observation is
that one can realise this lattice by definining $D\left( x_{1}\right) \vee
\ldots \vee D\left( x_{n}\right) $ as the radical of the ideal generated by $%
x_{1},\ldots ,x_{n}$. In particular the least element $0$ of the lattice is
turned into the nilradical $\sqrt{0}$, and $D\left( u\right) =0$ is
interpreted as $u\in \sqrt{0}$. The resulting proof \cite{per:spe} is fully
constructive, and works without any of the decidability assumptions we will
have to make---and to eliminate eventually by a combination of the G\"{o}%
del--Gentzen and Dragalin--Friedman proof translations. In \cite{per:spe}
only the essence of this elimination method occurs, already within the proof
and at a lower level.

In our own constructive proof of NC we still follow the lines along which we
have deduced KL\ from OI, but since the hypothesis of NC is computationally
more informative than the one of KL, we can get by with much less:\ with FI\
in place of OI. For short, we pass from the top to the bottom side of the
following square:%
\[
\begin{array}{ccc}
\text{OI} & \rightarrow & \text{KL} \\ 
\downarrow &  & \downarrow \\ 
\text{FI} & \rightarrow & \text{NC}%
\end{array}%
\]%
Yet we have to make a move that in the first place may seem
nonconstructive:\ as we had to assume (a variant of) SPT before, we now
employ another type of a classically valid distinction-by-cases, which has
occurred in constructive and computable algebra \cite%
{lomqui:pmbook,min:bib,sto:comprings}. However, as we have hinted at above
and will sketch below (Section \ref{elim}), this use of fragments of the Law
of Excluded Middle can be eliminated by proof theory.

\subsection{Constructive Proofs by Induction}

\subsubsection{With the Proof Pattern}

To deduce NC from FI, let $X$ be the partial order that consists of the
radical ideals of the ideals generated by some of the nonconstant
coefficients of $f$ and $g$. In other words, an element $F$ of $X$ is of the
form $F=\sqrt{I}$ where $I=(D)$ is the ideal generated by a detachable
subset $D$ of the set $E$ of the nonconstant coefficients of $f$ and $g$:
that is, 
\[
E=\{a_{1},\ldots ,a_{n},b_{1},\ldots ,b_{m}\}. 
\]%
This $X$, ordered by inclusion, possesses $\bot =\sqrt{0}$ and $\top =\sqrt{%
(E)}$, corresponding to $D=\emptyset $ and $D=E$. We assume that $r\in F$ is
decidable for all $r\in E$ and $F\in X$; whence in particular the partial
order $X$ is finite in the sense coined before (recall that $\sqrt{I}%
\subseteq \sqrt{J}$ if and only if $I\subseteq \sqrt{J}$). Now define the
predicate $U$ on $X$ by%
\[
U(F)\equiv u\in F\,. 
\]%
Both $X$ and $U$ are sets in \textbf{CZF}. Again $U(\top )$, and $U$ is
meet-closed and monotone. Once more our goal is to show $U(\bot )$, and to
apply the proof pattern from Theorem \ref{PP} we prove that $U$ is good. Let 
$F\in X$. By our decidability assumption we can distinguish the following
two cases.

\emph{Case 1.}\ If $a_{i}\in F$ for all $i>0$, then $u\in F$ and thus $%
U\left( F\right) $.

\emph{Case 2.}\ If $a_{i}\notin F$ for some $i>0$, then by (\ref{invneu})
also $b_{j}\notin F$ for some $j>0$. In this case---following Gau\ss 's
trick again---we pick $i,j$ that are maximal of this kind, for which (where
in each sum $p+q=i+j$) {%
\begin{equation}
F\ni c_{i+j}=\underbrace{\sum_{q>j}a_{p}b_{q}}_{\in F}+a_{i}b_{j}+%
\underbrace{\sum_{p>i}a_{p}b_{q}}_{\in F}  \label{GaussNeu}
\end{equation}%
and thus }$a_{i}b_{j}\in F$. Pretty much as in the deduction of KL from OI,
one can now see that $F$ is reducible. In detail, let $F=\sqrt{I}$ where $%
I=(D)$ for a detachable subset $D$ of $E$, and set%
\begin{equation}
G=\sqrt{I+Ra_{i}}\,,\quad H=\sqrt{I+Rb_{j}}\,,  \label{children}
\end{equation}%
for which $G,H\in X$, and $F\subsetneqq G$ and $F\subsetneqq H$ according to
the particular choice of $i$ and $j$. Moreover, $F=\sqrt{F}$ since $F$ is a
radical ideal; and $\sqrt{F}=G\cap H$ by Lemma \ref{key} and because $%
a_{i}b_{j}\in F$. Hence $F=G\cap H$ is the required decomposition of $F$.

\subsubsection{An Alternative Proof}

We now give an alternative deduction of NC\ from FI in which the exponents
are moved from $X$ to $U$. This allows for a conceptually simpler $X$, and
for a better understanding of the corresponding tree and algorithm (see
below). However we can no longer follow the proof pattern encapsulated in
Theorem \ref{PP}, because Lemma \ref{key} fails once the radicals are
removed. The modified predicate can still be proved to be progressive, and
induction is possible.

Here let $X$ be the partial order that consists of \emph{all} the ideals
generated by some of the nonconstant coefficients of $f$ and $g$: that is,
an element of $X$ is of the form $I=(D)$ where $D$ is a detachable subset of 
$E$ as before. Again, $X$ has finitely many elements, and we may assume that 
$\subseteq $ is a decidable relation on $X$, which is to say that%
\begin{equation}
\forall r\in E\,\forall F\in X\,\left( r\in F\vee r\notin F\right) \,.
\label{dec}
\end{equation}%
In all, $X$ ordered by inclusion is a finite partial order. Now we define
the predicate $U$ on $X$ in a slightly different way by%
\[
U(I)\equiv \exists e\in \mathbb{N}\,(u^{e}\in I)\,.
\]%
Once more both $X$ and $U$ are sets in \textbf{CZF}, and $U$ is monotone. To
prove $U\left( 0\right) $ by induction, or equivalently that $U\left(
I\right) $ for all $I\in X$, let $I\in X$. As before yet with $I$ in place
of $F$, we distinguish two cases.

\emph{Case 1.}\ If $a_{i}\in I$ for all $i>0$, then $u\in I$, and $e=1$
witnesses $U\left( I\right) $.

\emph{Case 2.}\ If $a_{i}\notin I$ for some $i>0$, then by (\ref{invneu})
also $b_{j}\notin I$ for some $j>0$. Pick $i,j$ that are maximal of this
kind. As before, still with $I\ $in place of $F$, one can show that $%
a_{i}b_{j}\in I$. Set 
\begin{equation}
K=I+Ra_{i}\,,\quad L=I+Rb_{j}\,  \label{childrensimple}
\end{equation}%
Now $K,L\in X$, and $I\subsetneqq K$ and $I\subsetneqq L$. By induction, $%
U\left( K\right) $ and $U\left( L\right) $: that is, there are $k,\ell \in 
\mathbb{N}$ such that $u^{k}\in K$ and $u^{\ell }\in L$. Hence $u^{k}u^{\ell
}\in I+Ra_{i}b_{j}$ (see the proof of Lemma \ref{key}), and thus $u^{k+\ell
}\in I$ because $a_{i}b_{j}\in I$;\ so $e=k+\ell $ witnesses $U\left(
I\right) $.

\subsection{Tree and Algorithm}

\subsubsection{Growing a Tree}

It is well-known how a tree can be grown along a proof by induction. We next
instantiate this method for the preceding proof, the notations and
hypotheses of which we adopt. In parallel to creating the nodes, we label
them by elements of $X$. To start the construction, we label the root by $0$%
. If a node $N$ labelled by $I\in X$ has just been constructed, then we
proceed according to the distinction-by-cases made during the proof, as
follows:

\emph{Case 1.} Declare $N$ to be a leaf.

\emph{Case 2.}\ Endow $N$ with two children labelled by $K$ and $L$ as in (%
\ref{childrensimple}).\newline
We thus get a full binary tree: every node either is a leaf or else is a
parent with exactly two children. Moreover the labelling is strictly
increasing:\ if a parent is labelled by $I$,$\ $and any one of its children
by $J$, then $I\subsetneqq J$. In particular, the tree is finite.

By construction, the label $I$ of a node $N$ satisfies $U$ whenever either $%
N $ is a leaf or else $N$ is a parent both children of which have labels
satisfying $U\smallskip $. In fact, in Case 1 we have $U(I)$ anyway; in Case
2 if $U(K)$ and $U(L)$, then $U(I)$ as shown in the proof. Climbing down
from the leaves to the root---that is, doing induction on the height of
a node, i.e.~its distance from the nearest leaf---one can thus show $U\left(
I\right) $ for every $I$ that occurs as the label of a node. In particular,
the label $0$ of the root satisfies $U$: that is, $U\left( I\right) $ for all $%
I\in X$.

\subsubsection{About Size}

To get an idea of the size of the tree we review its construction in terms
of the generators of the labels. First, the label $0$ of the root is
generated by the empty set $\emptyset $. Secondly, the label of a child is
obtained by adding a single element to the generators of the label $I$ of
the parent: an element $a_{i}$ of $\{a_{1},\ldots ,a_{n}\}\setminus I$ for
the one child and an element $b_{j}$ of $\{b_{1},\ldots ,b_{m}\}\setminus I$
for the other child, where both $i$ and $j$ are maximal among the remaining
indices. Thirdly, a node is a leaf whenever either all the $a_{1},\ldots
,a_{n}$ or equivalently all the $b_{1},\ldots ,b_{m}$ belong to the
generators of the label.

This said, what are the extremal lengths of the paths from the root to the
leaves? The height of the tree, i.e.~the length of the longest path, is at
most $n+m-1$. In fact, the longest paths have to be taken whenever for the
choice of new generators one keeps switching between the $a_{1},\ldots
,a_{n} $ and the $b_{1},\ldots ,b_{m}$. This is the case, for example, if
one adds first $a_{n}$, secondly $b_{m}$, thirdly $a_{n-1}$, next $b_{m-1}$,
and so on. From the root this requires adding all the $a_{2},\ldots ,a_{n}$
and all the $b_{2},\ldots ,b_{m}$, and thus possibly $n-1+m-1$ nodes, before
one arrives at a leaf by eventually adding either $a_{1}$ or $b_{1}$.

However there are shorter paths, which have length\ $\leq \min \{n,m\}$: the
path along which only the $a_{1},\ldots ,a_{n}$ (respectively, only the $%
b_{1},\ldots ,b_{m}$) are successively added to the generators has length $%
\leq n$ (respectively, length $\leq m$). Even shorter paths are possible
whenever the $a_{1},\ldots ,a_{n}$ and $b_{1},\ldots ,b_{m}$ fulfil
additional conditions; some of these coefficients may indeed be equal or
otherwise related in an appropriate way. In general however the tree is
uniform in the given data. We henceforth assume the generic situation in
which the $a_{0},\ldots ,a_{n}$ and $b_{0},\ldots ,b_{m}$ do not satisfy any
further algebraic dependence relation apart from (\ref{explizit}), and
accordingly can be seen as indeterminate coefficients \cite[p.~82]%
{lomqui:pmbook} only subject to (\ref{explizit}).

\subsubsection{Removing Redundancy}

The tree is repetitive inasmuch as some subtrees occur several times. To
remove this redundancy, one can identify all subtrees of the same form, and
rearrange the arrows accordingly. One thus transforms the tree into a simple
acyclic digraph with the source and the sinks coming from the root and the
leaves, respectively:%
\[
\!\!\!
\begin{array}{ccccccccc}
0 & \!\!\!\rightarrow\!\!\!  & (b_{m}) & \!\!\!\rightarrow\!\!\!  & \ldots  & \!\!\!\rightarrow\!\!\!  & 
(b_{2},\ldots ,b_{m}) & \!\!\!\rightarrow\!\!\!  & (b_{1},\ldots ,b_{m}) \\ 
\downarrow  &  & \downarrow  &  &  &  & \downarrow  &  &  \\ 
(a_{n}) & \!\!\!\rightarrow\!\!\!  & (a_{n},b_{m}) & \!\!\!\rightarrow\!\!\!  & \ldots  & 
\!\!\!\rightarrow\!\!\!  & (a_{n},b_{2},\ldots ,b_{m}) & \!\!\!\rightarrow\!\!\!  & 
(a_{n},b_{1},\ldots ,b_{m}) \\ 
\downarrow  &  & \downarrow  &  &  &  & \downarrow  &  &  \\ 
\vdots  &  & \vdots  &  &  &  & \vdots  &  & \vdots  \\ 
\downarrow  &  & \downarrow  &  &  &  & \downarrow  &  &  \\ 
(a_{2},\ldots ,a_{n}) & \!\!\!\rightarrow\!\!\!  & (a_{2},\ldots ,a_{n},b_{m}) & 
\!\!\!\rightarrow\!\!\!  & \ldots  & \!\!\!\rightarrow\!\!\!  & (a_{2},\ldots ,a_{n},b_{2},\ldots
,b_{m}) & \!\!\!\rightarrow\!\!\!  & (a_{2},\ldots ,a_{n},b_{1},\ldots ,b_{m}) \\ 
\downarrow  &  & \downarrow  &  &  &  & \downarrow  &  &  \\ 
(a_{1},\ldots ,a_{n}) &  & (a_{1},\ldots ,a_{n},b_{m}) &  & \ldots  &  & 
(a_{1},\ldots ,a_{n},b_{2},\ldots ,b_{m}) &  & 
\end{array}%
\]%
While the only source is in the top left corner, the $n+m$ sinks form the
rightmost column and the bottom row. There are $nm+n+m$ vertices; and if we
removed the sinks, then we would get a square grid graph with $n$ rows and $m
$ columns.

\subsubsection{Computing Witnesses}

In the foregoing we have proved constructively, with FI, that $u^{e}=0$ for
some $e\in \mathbb{N}$. Hence the tree grown alongside the induction encodes
an algorithm to compute a witness for this existential statement:\ that is,
an exponent $e$ under which $u^{e}=0$. This algorithm terminates since, as
we have observed before, the tree is finite.

Proof, tree, and algorithm are furthermore independent of the choice of $u$
among the $a_{1},\ldots ,a_{n}$; and thus only depend on $n$ and $m$. In
fact the $e$ produced by the algorithm works for each of the $u$ from the $%
a_{1},\ldots ,a_{n}$, though---as in the example below---a smaller exponent
may suffice for some. One could also have shortened proof, tree, and
algorithm by stopping already when the given $u$ belongs to the ideal $I$
under consideration; for the sake of uniformity we have disregarded this
option.

We now look back to see where the exponents come from and how they grow
during the course of the algorithm.\footnote{%
The author is indebted to Ulrich Berger for prompting these investigations.}
For any node $N$ labelled by $I\in X$, we say that $e$\emph{\ witnesses }$%
U\left( I\right) $ whenever $u^{e}\in I$. If $N\ $is a leaf (Case 1), then 1
witnesses $U\left( I\right) $. If $N$ is a parent (Case 2) labelled by $I$,
with children labelled by $K$ and $L$, and $k$ and $\ell $ witness $U\left(
K\right) $ and $U\left( L\right) $, respectively, then $k+\ell $ witnesses $%
U\left( I\right) $, as we see from the proof. Hence $U\left( 0\right) $ is
witnessed by the number of leaves, which in turn is bounded by $2^{n+m-1}$
(recall that the height of the tree is at most $n+m-1$).

To get a sharper bound, one may switch to the digraph, this time labelled by
the exponents and with reversed arrows:%
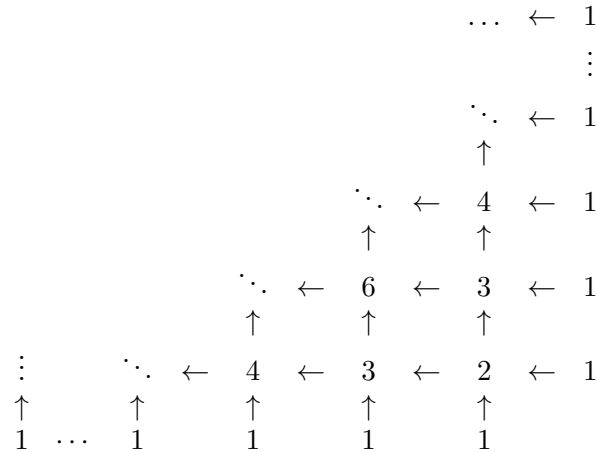
\[
\begin{array}{ccccccccccc}
&  &  &  &  &  &  &  & \ldots  & \leftarrow  & 1 \\ 
&  &  &  &  &  &  &  &  &  & \vdots  \\ 
&  &  &  &  &  &  &  & \ddots  & \leftarrow  & 1 \\ 
&  &  &  &  &  &  &  & \uparrow  &  &  \\ 
&  &  &  &  &  & \ddots  & \leftarrow  & 4 & \leftarrow  & 1 \\ 
&  &  &  &  &  & \uparrow  &  & \uparrow  &  &  \\ 
&  &  &  & \ddots  & \leftarrow  & 6 & \leftarrow  & 3 & \leftarrow  & 1 \\ 
&  &  &  & \uparrow  &  & \uparrow  &  & \uparrow  &  &  \\ 
\vdots  &  & \ddots  & \leftarrow  & 4 & \leftarrow  & 3 & \leftarrow  & 2 & 
\leftarrow  & 1 \\ 
\uparrow  &  & \uparrow  &  & \uparrow  &  & \uparrow  &  & \uparrow  &  & 
\\ 
1 & \cdots  & 1 &  & 1 &  & 1 &  & 1 &  & 
\end{array}%
\]%
This in fact is a finite fragment of Pascal's triangle; whence standard
theory can be used to calculate the number in the top left corner: that is,
the $e$ under which $u^{e}=0$.

Note in this context that the construction of the tree and thus the design
of the algorithm are not affected by the relations expressing the hypothesis 
$fg=1$. This information is only used for proving that the algorithm meets
its specification:\ that is, $u^{e}=0$ whenever $e$ is the exponent present
at the root. More specifically, (\ref{implizit}) and (\ref{explizit}) are
invoked for proving (\ref{invneu}) and (\ref{GaussNeu}), and thus for
proving that in Case 2 above the certificate for $e$ witnessing $U(I)$ is
bestowed from the children to the parent.

\subsubsection{A Concrete Example}

Let $n=2$ and $m=1$, and set%
\[
G=(b_{1})\,,~F=(a_{2})\,,~L=(a_{2},b_{1})\,,~K=(a_{1},a_{2})\,. 
\]
The digraphs labelled by finitely generated ideals and exponents are as
follows: 
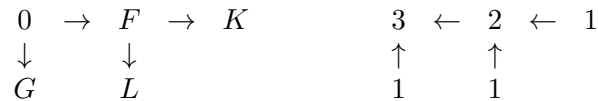
\[
\begin{array}{ccccc}
0 & \rightarrow & F & \rightarrow & K \\ 
\downarrow &  & \downarrow &  &  \\ 
G &  & L &  & 
\end{array}%
\quad \quad \quad \quad 
\begin{array}{ccccc}
3 & \leftarrow & 2 & \leftarrow & 1 \\ 
\uparrow &  & \uparrow &  &  \\ 
1 &  & 1 &  & 
\end{array}%
\]%
In particular, $e=3$ is the output exponent under which $u^{e}=0$ for any
choice of $u$.

Now, for the sake of simplicity, assume that $a_{0}=1$ and $b_{0}=1$. Apart
from $a_{0}b_{0}=1$, which in this case is trivial, (\ref{explizit}) then
contains the following information:%
\[
a_{1}+b_{1}=0\,,\quad a_{1}b_{1}+a_{2}=0\,,\quad a_{2}b_{1}=0\,. 
\]%
With these equations at hand, the certificates for $u^{e}\in I$ where $u\in
\{a_{1},a_{2}\}$ and $I\in \{K,L,F,G,0\}$ are achieved as follows:%
\[
\begin{array}{ll}
\text{Case }u=a_{1}\text{.} & \text{Case }u=a_{2}\text{.} \\ 
a_{1}\in K & a_{2}\in K \\ 
a_{1}=-b_{1}\in L & a_{2}\in L \\ 
a_{1}^{2}=-a_{1}b_{1}=a_{2}\in F & a_{2}\in F \\ 
a_{1}=-b_{1}\in G & a_{2}=-a_{1}b_{1}\in G \\ 
a_{1}^{3}=a_{1}^{2}a_{1}=-a_{2}b_{1}=0 & a_{2}^{2}=-a_{2}a_{1}b_{1}=0%
\end{array}%
\]%
Note that $e=2$ suffices for $u=a_{2}$, whereas $e=3$ is required for $%
u=a_{1}$.

Finally, let $R=\mathbb{Z}/(8)$, and set $a_{1}=2$, $a_{2}=4$, $b_{1}=6$. In
this case, 
\[
f=4T^{2}+2T+1\,,\quad g=6T+1\,\,, 
\]%
for which indeed, as we are doing integer arithmetic modulo 8, 
\[
fg=24T^{3}+16T^{2}+8T+1=1\,. 
\]%
Here $e=2$ suffices for $u=a_{2}=4$, whereas $e=3$ is required for $%
u=a_{1}=2 $.

\subsubsection{A Representation}

For any implementation on a computer, a more concrete representation of the
elements of $X$ may be required:\ that is, of the ideals generated by a
detachable subset of $E$. Especially in view of the independence assumption
made above, a natural choice is to represent any element $I$ of $X$ by a
pair $\left( \lambda ,\mu \right) $ of increasing binary lists $\lambda
=\left( \lambda _{1}\ldots \lambda _{n})\text{ and }\mu =(\mu _{1}\ldots \mu
_{m}\right) $ of length $n$ and $m$. Here $\lambda _{i}=1$ and $\mu _{j}=1$
indicate that $a_{i}$ and $b_{j}$, respectively, belong to the generators of 
$I$. The inclusion order on $X$ is then represented by the pointwise order
of binary sequences.

With this representation at hand the tree can be described as follows. The
root is labelled by $(0^{n},0^{m})$. A generic node is labelled by $%
(0^{k}1^{\ell },0^{p}1^{q})$ where $k+\ell =n$ and $p+q=m$; and has two
children with labels $(0^{k-1}1^{\ell +1},0^{p}1^{q})$ and $(0^{k}1^{\ell
},0^{p-1}1^{q+1})$, unless either $k=n$ or $\ell =m$, in which case this
node is a leaf. If we write $\overline{0}$ and $\overline{1}$ for finite
lists $0\ldots 0$ and $1\ldots 1$, respectively, of variable but appropriate
lengths, then the corresponding digraph is as follows:%
\[
\begin{array}{ccccccccc}
(\overline{0},\overline{0}) & \rightarrow  & (\overline{0},\overline{0}1) & 
\rightarrow  & \ldots  & \rightarrow  & (\overline{0},0\overline{1}) & 
\rightarrow  & (\overline{0},\overline{1}) \\ 
\downarrow  &  & \downarrow  &  &  &  & \downarrow  &  &  \\ 
(\overline{0}1,\overline{0}) & \rightarrow  & (\overline{0}1,\overline{0}1)
& \rightarrow  & \ldots  & \rightarrow  & (\overline{0}1,0\overline{1}) & 
\rightarrow  & (\overline{0}1,\overline{1}) \\ 
\downarrow  &  & \downarrow  &  &  &  & \downarrow  &  &  \\ 
\vdots  &  & \vdots  &  &  &  & \vdots  &  & \vdots  \\ 
\downarrow  &  & \downarrow  &  &  &  & \downarrow  &  &  \\ 
(0\overline{1},\overline{0}) & \rightarrow  & (0\overline{1},\overline{0}1)
& \rightarrow  & \ldots  & \rightarrow  & (0\overline{1},0\overline{1}) & 
\rightarrow  & (0\overline{1},\overline{1}) \\ 
\downarrow  &  & \downarrow  &  &  &  & \downarrow  &  &  \\ 
(\overline{1},\overline{0}) &  & (\overline{1},\overline{0}1) &  & \ldots  & 
& (\overline{1},0\overline{1}) &  & 
\end{array}%
\]

\subsubsection{Elimination of Decidability\label{elim}}

We roughly sketch, as promised before, how we can eliminate the classically
valid decidability assumptions, such as (\ref{dec}), used in the
constructive proofs. Those assumptions form a finite set $\Delta $ of
instances of the Law of Excluded Middle. Let $\Gamma $ consist of the
finitely many equations listed in (\ref{explizit});\ and let $\vdash _{i}$
and $\vdash _{c}$ stand for deducibility with intuitionistic and classical
logic, respectively. If we neglect technical details, then with our
constructive proof above---of NC with FI---we have established $\Gamma
,\Delta \vdash _{i}C$ where%
\[
C\equiv \exists e\,(u^{e}=0)\,.
\]%
Since, clearly, $\vdash _{c}\Delta $, we thus have $\Gamma \vdash _{c}C$.
Now, since both $C$ and the elements of $\Gamma $ are of a sufficiently
simple logical form, e.g.~geometric formulas, by proof--theoretic techniques
such as syntactic versions of Barr's Theorem---see e.g.~\cite%
{ish:translation,negri:barr,palm:real}---we arrive at $\Gamma \vdash _{i}C$.
The logical form of the formulas in $\Delta $ is irrelevant for 
this argument; and a similar method \cite{val:phd} applies if $C$ is seen 
as an infinite disjunction rather than as an existential formula. 

One possible road from $\Gamma \vdash _{c}C$ to $\Gamma \vdash _{i}C$ is via
the generalisation (see e.g.~\cite{ish:translation,palm:real}) of the G\"{o}%
del--Gentzen negative translation for which, in the spirit of the
Dragalin--Friedman $A$--translation, the falsum $\bot $ is replaced by an
arbitrary formula $A$. This $A$ typically is the conclusion of the deduction
under consideration.\footnote{%
Thierry Coquand has kindly pointed us toward this method. As we have learned
from Christoph-Simon Senjak, the proof translation at use is related---via
the Curry-Howard isomorphism---to the continuation-passing style in
programming.} In particular, one sets $\bot ^{A}\equiv A$, and an atomic
formula $B\not\equiv \bot $ is assigned to%
\[
B^{A}\equiv \left( B\rightarrow A\right) \rightarrow A\,;
\]%
the existential quantifier moreover is translated as follows:%
\[
\left( \exists x\,B\right) ^{A}\equiv \forall x\left( B^{A}\rightarrow
A\right) \rightarrow A\,.
\]%
Hence if $B\not\equiv \bot $ is atomic, then%
\[
\vdash _{i}\left( B^{A}\rightarrow A\right) \leftrightarrow \left(
B\rightarrow A\right) 
\]%
and thus%
\begin{equation}
\vdash _{i}\left( \exists x\,B\right) ^{A}\leftrightarrow \left( \exists
xB\rightarrow A\right) \rightarrow A\,  \label{trans}
\end{equation}%
provided that the variable $x$ does not occur freely within the formula $A$.

With this translation, and for $\Gamma $ and $C$ as above, one can prove
that $\Gamma \vdash _{c}C$ implies $\Gamma ^{A}\vdash _{i}C^{A}$. In view of
the simple form of the elements of $\Gamma $, we further have $\Gamma \vdash
_{i}\Gamma ^{A}$. Hence $\Gamma \vdash _{c}C$ implies $\Gamma \vdash
_{i}C^{A}$, which in the specific case $A\equiv C$ yields $\Gamma \vdash
_{i}C$. In fact, we have $\vdash _{i}C^{C}\leftrightarrow \left( C\rightarrow
C\right) \rightarrow C$ by (\ref{trans}), and thus $\vdash
_{i}C^{C}\leftrightarrow C$.

\section{Conclusion}

Our choices of the partial orders $X$ and the predicates $U$ were crucial to
make induction work. In particular, the objects $X$ from the constructive
proofs are finite partial orders, as required for FI, and, by the way, are
sets in the sense of \textbf{CZF}. Moreover, we thus have eventually kept
close to the data of the given problem: that is, the coefficients $a_{i}$
and $b_{j}$ of the polynomials $f$ and $g$. We could have done so much
earlier, and perhaps more efficiently: by the method of indeterminate
coefficients \cite[p.~82]{lomqui:pmbook}. This would have meant to pass from
the arbitrary given $R$ to the ring%
\[
R_{0}=\mathbb{Z}[a_{0},\ldots ,a_{n},b_{0},\ldots
,b_{m}]/(a_{0}b_{0}-1,a_{0}b_{1}+a_{1}b_{0},\ldots ,a_{n}b_{m}\,)\text{%
\thinspace }.
\]%
with generators $a_{0},\ldots ,a_{n},b_{0},\ldots ,b_{m}$ and relations (\ref%
{explizit}). This $R_{0}$ indeed encodes all the data and information we
would have needed for phrasing and proving NC. For its simple structure,
moreover, $R_{0}$ is Noetherian \cite{jacobsson:standard};\ and SPT holds
constructively for $R_{0}$ since this is a fully Lasker--Noether ring \cite%
{perdry:noether}. Hence LN for $R_{0}$ is constructively provable \cite%
{perdry:noether}, and KL for $R_{0}$ follows without any talk of OI, let
alone of ZL.

The universal quantification required for KL, over all possible prime ideals
of $R$, could then be replaced by the more manageable one over the finitely
many finitely generated prime ideals of $R_{0}$ as produced by LN. Modulo
each of the latter prime ideals we could have followed Gau\ss 's argument
for the case of a domain, yet replacing---as we have done anyway---the proof
by contradiction by an appropriate distinction-by-cases. In all, we would
have got a perfectly constructive proof of NC. This avenue, however, might
not have forced us to seek an invocation of FI, and thus to keep close to
the given data. In particular, we would not have made explicit a tree and an
algorithm as simple as they have resulted from the constructive proofs with
FI.

With hindsight, the proof pattern coined with Theorem \ref{PP}, including
the crucial notion of reducibility, stands already behind many a post-war
textbook proof of the Lasker--Noether theorem \cite{AMD,north:ideal}, and
more implicitly behind Krull's proof \cite{krull:ideal}. It is not yet clear
however whether any constructive \textquotedblleft
unwinding\textquotedblright\ of this type of classical proof can be brought
under that pattern. We have anyway seen how the pattern can be applied to
NC, a lemma in polynomial algebra, and yet another application of the
pattern has proved possible in the area of inversion problems for Banach
algebras \cite{CiE2012}. A further case study will be undertaken about Gau%
\ss 's lemma \textquotedblleft the product of two primitive polynomials is
primitive\textquotedblright , which leads over its generalisation, ascribed
to Joyal, to the so-called Dedekind Prague Theorem \cite%
{ban:pol,coq:prague,lom:int,coq:valspace}.

We have conceived the constructive proofs of NC along the lines of the
classical proof with Gau\ss 's trick that we have recalled earlier. In
particular, (\ref{invneu}) and (\ref{GaussNeu}) are nothing but (\ref{invorg}%
) and (\ref{GaussOrg}), respectively, with \textquotedblleft $=0$%
\textquotedblright\ replaced by \textquotedblleft $\in F$\textquotedblright
. Now if \textquotedblleft $=0$\textquotedblright\ is considered within a
domain, as in the classical proof, then it actually corresponds to
\textquotedblleft $\in P$\textquotedblright\ where $P$ is a prime ideal of
an arbitrary ring. Hence the move from \textquotedblleft $=0$%
\textquotedblright\ to \textquotedblleft $\in F$\textquotedblright\ is in
accordance with a paradigm that goes back to the so-called D5 philosophy of
dynamic evaluation in computer algebra\ \cite{duval:about}: to handle the
prime ideals $P$ by way of their incomplete specifications $F$ \ \cite%
{cos:dyn,lombardi:krull,lom:int}, which are (radicals of) finitely generated
ideals but not necessarily prime.

In all, we have carried out a case study for a further potentially
systematic way to gain finite methods from ideal objects in algebra. As
discussed above (Section \ref{discuss}), our main competitor \cite{per:spe}
has studied the same case but with a different method taken from point-free
topology. We thus have started yet another attempt to make constructive
sense of the notion of prime ideals, which \textquotedblleft \ldots\ play a
central role in the theory of commutative rings\textquotedblright\ \cite[p.~1%
]{kap:comrgs}. In constructive algebra the notion of prime ideals has seen a
revival \cite{bri:pri} after it was considered problematic in general:
\textquotedblleft If an ideal $P$ in a commutative ring is not detachable,
it is is not clear just what it should mean for $P$ to be
prime\textquotedblright\ \cite[p.~77]{min:bib}.

Our work may further give evidence for the practicability of the recent
proposals of a controlled use of ideal objects in constructive mathematics 
\cite{sam:dyn,sam:realideal} on the basis of a two-level foundations with
forget-restore option \cite{maisam:minimalist,maietti:minimalist}. Last but
not least, we have put some mathematical flesh on Bell's conjecture that ZL
is \textquotedblleft constructively neutral\textquotedblright\ \cite%
{bell:zornneutral}.

\section*{Acknowledgements}

Useful hints came from many participants of the 2011 Oberwolfach Workshop on
Proof Theory and Constructive Mathematics, such as Ulrich Berger, Thierry
Coquand, Erik Palmgren, and Per Martin-L\"{o}f. The author is further
grateful to many others---especially to Martin Hofmann, Henri Lombardi,
Davide Rinaldi, Pedro Francisco Valencia Vizca{\'i}no, and Olov Wilander---for 
stimulating discussions; to Jean-Claude Raoult, Bernhard Reus, and Fred Richman for
commenting on draft versions; to the anonymous referees for their constructive critique; 
and last but not least to Matthew Hendlass, who first asked for a proof pattern.

This paper was first written during a fellowship at the Isaac Newton
Institute for Mathematical Sciences, programme \textquotedblleft Semantics
and Syntax: A Legacy of Alan Turing\textquotedblright . It was revised later
during a visit to the University of Stockholm funded by 
the European Science Foundation Research Networking Programme
\textquotedblleft New frontiers of infinity: mathematical, philosophical,
and computational prospects\textquotedblright ; and during a visit to
Swansea University with a Computer Science Small Grant of the London
Mathematical Society.%

This line of research was started when the author had a Feodor Lynen
Research Fellowship for Experienced Researchers granted by the Alexander von
Humboldt Foundation from sources of the German Bundesministerium f\"{u}r
Bildung und Forschung; and when he was a visiting professor supported by a
grant from the Italian Istituto Nazionale di Alta Matematica---Gruppo
Nazionale per le Strutture Algebriche, Geometriche e le loro Applicazioni.
The author is particularly grateful to Andrea Cantini, Giovanni Sambin, and
various colleagues at Padua and Florence for their most welcoming
hospitality.

\bibliographystyle{plain}
\bibliography{besancon}

\begin{thebibliography}{10}

\bibitem{AczelInter}
Peter Aczel.
\newblock The type theoretic interpretation of constructive set theory.
\newblock In {\em Logic {C}olloquium '77 ({P}roc. {C}onf., {W}roc\l aw, 1977)},
  volume~96 of {\em Stud. Logic Foundations Math.}, pages 55--66.
  North-Holland, Amsterdam, 1978.

\bibitem{AczelChoice}
Peter Aczel.
\newblock The type theoretic interpretation of constructive set theory: choice
  principles.
\newblock In {\em The {L}. {E}. {J}. {B}rouwer {C}entenary {S}ymposium
  ({N}oordwijkerhout, 1981)}, volume 110 of {\em Stud. Logic Found. Math.},
  pages 1--40. North-Holland, Amsterdam, 1982.

\bibitem{AczelInd}
Peter Aczel.
\newblock The type theoretic interpretation of constructive set theory:
  inductive definitions.
\newblock In {\em Logic, methodology and philosophy of science, {VII}
  ({S}alzburg, 1983)}, volume 114 of {\em Stud. Logic Found. Math.}, pages
  17--49. North-Holland, Amsterdam, 1986.

\bibitem{AMD}
Michael~F. Atiyah and Ian~G. Macdonald.
\newblock {\em Introduction to Commutative Algebra}.
\newblock Addison-Wesley Publishing Co., 1969.

\bibitem{ban:pol}
B.~Banaschewski and J.~J.~C. Vermeulen.
\newblock Polynomials and radical ideals.
\newblock {\em J. Pure Appl. Algebra}, 113(3):219--227, 1996.

\bibitem{ban:rad}
Bernhard Banaschewski.
\newblock Radical ideals and coherent frames.
\newblock {\em Comment. Math. Univ. Carolin.}, 37(2):349--370, 1996.

\bibitem{bell:zornneutral}
John~L. Bell.
\newblock Zorn's lemma and complete {B}oolean algebras in intuitionistic type
  theories.
\newblock {\em J.~Symbolic Logic}, 62(4):1265--1279, 1997.

\bibitem{berger:open}
Ulrich Berger.
\newblock A computational interpretation of open induction.
\newblock In F.~Titsworth, editor, {\em Proceedings of the Ninetenth Annual
  IEEE Symposium on Logic in Computer Science}, pages 326--334. IEEE Computer
  Society, 2004.

\bibitem{bri:pri}
Douglas~S. Bridges.
\newblock {Prime and maximal ideals in constructive ring theory.}
\newblock {\em Commun.~Algebra}, 29:2787--2803, 2001.

\bibitem{Cohn:Univ}
Paul~M. Cohn.
\newblock {\em Universal Algebra}.
\newblock Harper \& Row Publishers, New York, 1965.

\bibitem{coq:combinatorics}
Thierry Coquand.
\newblock Constructive topology and combinatorics.
\newblock In {\em Constructivity in computer science ({S}an {A}ntonio, {TX},
  1991)}, volume 613 of {\em Lecture Notes in Comput. Sci.}, pages 159--164.
  Springer, Berlin, 1992.

\bibitem{coq:valspace}
Thierry Coquand.
\newblock Space of valuations.
\newblock {\em Ann.~Pure Appl.~Logic}, 157:97--109, 2009.

\bibitem{coq:logicalgebra}
Thierry Coquand and Henri Lombardi.
\newblock A logical approach to abstract algebra.
\newblock {\em Math.~Struct.~in Comput.~Science}, 16:885--900, 2006.

\bibitem{coqper:groebner}
Thierry Coquand and Henrik Persson.
\newblock Gr\"obner bases in type theory.
\newblock In {\em Types for proofs and programs ({I}rsee, 1998)}, volume 1657
  of {\em Lecture Notes in Comput. Sci.}, pages 33--46. Springer, Berlin, 1999.

\bibitem{coq:prague}
Thierry Coquand and Henrik Persson.
\newblock Valuations and {D}edekind's {P}rague theorem.
\newblock {\em J.~Pure Appl.~Algebra}, 155(2--3):121--129, 2001.

\bibitem{cos:dyn}
Michel Coste, Henri Lombardi, and Marie-Fran\c{c}oise Roy.
\newblock {Dynamical method in algebra: Effective Nullstellens\"atze.}
\newblock {\em Ann.~Pure Appl.~Logic}, 111(3):203--256, 2001.

\bibitem{cro:fin}
Laura Crosilla and Peter Schuster.
\newblock {Finite Methods in Mathematical Practice}.
\newblock In G.~Link and M.~Detlefsen, editors, {\em Formalism and Beyond},
  Mathematical Logic. {Ontos}, Heusenstamm, 201x.

\bibitem{duval:about}
Jean~Della Dora, Claire Dicrescenzo, and Dominique Duval.
\newblock About a new method for computing in algebraic number fields.
\newblock In {\em European Conference on Computer Algebra (2)}, pages 289--290,
  1985.

\bibitem{friedman:foundations}
Harvey Friedman.
\newblock Set theoretic foundations for constructive analysis.
\newblock {\em Ann.~of Math.~(2)}, 105(1):1--28, 1977.

\bibitem{CiE2012}
Matthew Hendtlass and Peter Schuster.
\newblock A direct proof of {W}iener's theorem.
\newblock In S.~B. Cooper, A.~Dawar, and B.~L\"owe, editors, {\em How the World
  Computes. Turing Centenary Conference and {E}ighth {C}onference~on
  {C}omputability in {E}urope}, volume 7318 of {\em Lect.~Notes Comput.~Sci.},
  pages 294--303, {Berlin and Heidelberg}, 2012. {Springer}.
\newblock Proceedings, CiE 2012, {C}ambridge, UK, June 2012.

\bibitem{hub:ind}
Simon Huber and Peter Schuster.
\newblock Maximalprinzipien und {I}nduktionsbeweise.
\newblock Technical report, University of Leeds, 2013.
\newblock In preparation.

\bibitem{ish:translation}
Hajime Ishihara.
\newblock A note on the {G}\"odel-{G}entzen translation.
\newblock {\em MLQ Math. Log. Q.}, 46(1):135--137, 2000.

\bibitem{jacobsson:standard}
Carl Jacobsson and Clas L{\"o}fwall.
\newblock Standard bases for general coefficient rings and a new constructive
  proof of {H}ilbert's basis theorem.
\newblock {\em J.~Symb.~Comput.}, 12(3):337--372, 1991.

\bibitem{joh:sto}
Peter~T. Johnstone.
\newblock {\em {Stone {S}paces.}}
\newblock Number~{3} in {Cambridge Studies in Advanced Mathematics}. {Cambridge
  etc.: Cambridge University Press}, 1982.

\bibitem{joyal:zar}
Andr{\'e} Joyal.
\newblock Les th\'eoremes de {C}hevalley-{T}arski et remarques sur l'alg\`ebre
  constructive.
\newblock {\em Cah.~Topol.~G\'{e}om.~Diff\'{e}r.~Cat\'{e}g.}, 16:256--258,
  1976.

\bibitem{kap:comrgs}
Irving Kaplansky.
\newblock {\em Commutative Rings}.
\newblock The University of Chicago Press, Chicago and London, 1974.
\newblock Revised edition.

\bibitem{krull:ideal}
Wolfgang Krull.
\newblock {\em Idealtheorie}.
\newblock Ergebnisse der Mathematik und ihrer Grenzgebiete, vol.~4, no.~3.
  Springer, Berlin, 1935.

\bibitem{lombardi:krull}
Henri Lombardi.
\newblock Dimension de {K}rull, {N}ullstellens\"atze et \'evaluation dynamique.
\newblock {\em Math.~Zeitschrift}, 242:23--46, 2002.

\bibitem{lom:int}
Henri Lombardi.
\newblock Hidden constructions in abstract algebra. {I}. {I}ntegral dependance.
\newblock {\em J. Pure Appl. Algebra}, 167:259--267, 2002.

\bibitem{lombardi:algdyn}
Henri Lombardi.
\newblock Alg\`ebre dynamique, espaces topologiques sans points et programme de
  {H}ilbert.
\newblock {\em Ann.~Pure Appl.~Logic}, 137:256--290, 2006.

\bibitem{lomqui:pmbook}
Henri Lombardi and Claude Quitt\'e.
\newblock {\em Alg\`ebre commutative.~M\'ethodes constructives.~Modules
  projectifs de type fini.}
\newblock Calvage \& Mounet, Paris, 2012.

\bibitem{maietti:minimalist}
Maria~Emilia Maietti.
\newblock {A minimalist two-level foundation for constructive mathematics.}
\newblock {\em Ann. Pure Appl. Logic}, 160(3):319--354, 2009.

\bibitem{maisam:minimalist}
Maria~Emilia Maietti and Giovanni Sambin.
\newblock {Toward a minimalist foundation for constructive mathematics}.
\newblock In L.~Crosilla and P.~Schuster, editors, {\em From {S}ets and {T}ypes
  to {T}opology and {A}nalysis}, volume~48 of {\em {O}xford {L}ogic {G}uides},
  pages 91--114. {Oxford: Oxford University Press}, 2005.

\bibitem{ITT}
Per Martin-L{\"o}f.
\newblock {\em Intuitionistic type theory}, volume~1 of {\em Studies in Proof
  Theory. Lecture Notes}.
\newblock Bibliopolis, Naples, 1984.
\newblock Notes by Giovanni Sambin.

\bibitem{min:bib}
Ray Mines, Fred Richman, and Wim Ruitenburg.
\newblock {\em {A Course in Constructive Algebra}}.
\newblock Springer, New York, 1988.
\newblock Universitext.

\bibitem{negri:barr}
Sara Negri.
\newblock Contraction-free sequent calculi for geometric theories with an
  application to {B}arr's theorem.
\newblock {\em Arch. Math. Logic}, 42(4):389--401, 2003.

\bibitem{north:ideal}
Douglas~G. Northcott.
\newblock {\em Ideal {T}heory}.
\newblock Cambridge University Press, 1953.

\bibitem{palm:real}
Erik Palmgren.
\newblock An intuitionistic axiomatisation of real closed fields.
\newblock {\em MLQ Math. Log. Q.}, 48(2):297--299, 2002.

\bibitem{perdry:noether}
Herv\'e Perdry.
\newblock Strongly {N}oetherian rings and constructive ideal theory.
\newblock {\em J.~Symb.~Comput.}, 37(4):511--535, 2004.

\bibitem{perdry:noeord}
Herv\'e Perdry and Peter Schuster.
\newblock {N}oetherian orders.
\newblock {\em Math.~Structures Comput.~Sci.}, 21:111--124, 2011.

\bibitem{per:spe}
Henrik Persson.
\newblock An application of the constructive spectrum of a ring.
\newblock In {\em Type Theory and the Integrated Logic of Programs}. Chalmers
  University and University of G\"{o}teborg, 1999.
\newblock PhD thesis.

\bibitem{raoult:open}
Jean-Claude Raoult.
\newblock Proving open properties by induction.
\newblock {\em Inform.~Process.~Lett.}, 29(1):19--23, 1988.

\bibitem{rich:trivial}
Fred Richman.
\newblock Nontrivial uses of trivial rings.
\newblock {\em Proc.~Amer.~Math.~Soc.}, 103(4):1012--1014, 1988.

\bibitem{sam:ifs}
Giovanni Sambin.
\newblock {Intuitionistic formal spaces---a first communication}.
\newblock In {\em Mathematical Logic and its Applications, Proc. Adv. Internat.
  Summer School Conf., Druzhba, Bulgaria, 1986}, pages 187--204. {Plenum},
  1987.

\bibitem{sam:dyn}
Giovanni Sambin.
\newblock Steps towards a dynamic constructivism.
\newblock In P.~G\"{a}rdenfors et~al., editor, {\em In the {S}cope of {L}ogic,
  {M}ethodology and {P}hilosophy of {S}cience}, volume 315 of {\em Synthese
  Library}, pages 263--286, Dordrecht, 2002. Kluwer.
\newblock 11th {I}nternational {C}ongress of {L}ogic, {M}ethodology and
  {P}hilosophy of {S}cience. {K}rakow, {P}oland, {A}ugust 1999.

\bibitem{sam:som}
Giovanni Sambin.
\newblock Some points in formal topology.
\newblock {\em Theoret.~Comput.~Sci.}, 305(1-3):347--408, 2003.

\bibitem{sam:realideal}
Giovanni Sambin.
\newblock Real and ideal in constructive mathematics.
\newblock In {\em Epistemology versus ontology}, volume~27 of {\em Log.
  Epistemol. Unity Sci.}, pages 69--85. Springer, Dordrecht, 2012.

\bibitem{LICS2012}
Peter Schuster.
\newblock Induction in algebra: a first case study.
\newblock In {\em 2012 27th Annual ACM/IEEE Symposium on Logic in Computer
  Science}, pages 581--585. IEEE Computer Society Publications, 2012.
\newblock Proceedings, LICS 2012, Dubrovnik, Croatia, June 2012.

\bibitem{sto:comprings}
Viggo Stoltenberg-Hansen and John~V. Tucker.
\newblock Computable rings and fields.
\newblock In {\em Handbook of computability theory}, volume 140 of {\em Stud.
  Logic Found. Math.}, pages 363--447. North-Holland, Amsterdam, 1999.

\bibitem{val:phd}
Pedro Francisco~Valencia Vizca{\'i}no.
\newblock {\em Some Uses of Cut Elimination}.
\newblock Phd thesis, University of Leeds, 2013.

\bibitem{yengui:maximal}
Ihsen Yengui.
\newblock Making the use of maximal ideals constructive.
\newblock {\em Theoret.~Comput.~Sci.}, 392:174--178, 2008.

\end{thebibliography}

\end{document}